\documentclass[prl,reprint,aps,superscriptaddress,showpacs,showkeys,twocolumn,
	letterpaper]{revtex4-1}
\usepackage{amsmath}
\usepackage[breaklinks]{hyperref}
\usepackage{hyperref}
\usepackage{gensymb}
\usepackage {graphicx}
\usepackage{perpage}

\usepackage{multirow}
\newcommand {\nc} {\newcommand}
\newcommand {\rn} {\renewcommand}

\nc{\bittot}{Bi$_2$Sr$_2$CaCu$_2$O$_{8+\delta}$}
\nc{\capa}{\mathcal{C}_n}
\nc{\caparison}{\mathcal{C}_n(\vec{k}, \omega)}
\nc{\cuotwo}{CuO$_2$}
\nc{\ef}{E_F}
\nc{\ek}{\varepsilon(\kvec)}
\nc{\hightc}{high T$_c$}
\rn{\Im}{\mathrm{Im}\,}
\nc{\kf}{k_F}
\nc{\kvec}{\vec{k}}
\nc{\lsco}{La$_{2-x}$Sr$_x$CuO$_4$}
\nc{\pecfl}{pECFL}
\nc{\pecmd}{MD-pECFL}
\nc{\pecmi}{MI-pECFL}
\nc{\secfl}{sECFL}
\nc{\tc}{T$_c$}
\nc{\tj}{$t$-$J$}

\begin{document}

\title{Phenomenological model for the normal state ARPES line shapes of high temperature superconductors}
\author{Kazue Matsuyama}
\affiliation{Department of Physics, University of California, Santa Cruz, CA 95064}
\author{G.-H. Gweon}
\email{corresponding author, gweon@ucsc.edu.}
\affiliation{Department of Physics, University of California, Santa Cruz, CA 95064}

\date{\today}
\begin{abstract} 
Providing a full theoretical description of the single particle spectral function observed for high temperature superconductors in the normal state is an important goal, yet unrealized.  Here, we present a phenomenological model approaching towards this goal.  The model results from implementing key phenomenological improvement in the so-called extremely correlated Fermi liquid (ECFL) model.  The model successfully describes the dichotomy of the spectral function as functions of momentum and energy and fits data for different materials (\bittot{} and \lsco), with an identical set of intrinsic parameters.  The current analysis goes well beyond the prevalent analysis of the spectral function as a function of momentum alone.
\end{abstract} 


\maketitle

In the sudden approximation theory \cite{hedin__1969} of the angle resolved photo-electron spectroscopy (ARPES), photo-electron counts, $I(\kvec,\omega)$, recorded as a function of momentum ($\vec{k}$) and energy ($\omega$) \footnote{Throughout this Letter, $\hbar = 1$ by convention.} are given by
\begin{align}
	I(\kvec, \omega) &= |M_{if}|^2\, f(\omega)\, A (\kvec, \omega) \label{eq-sudden-approximation}
\end{align}
where $M_{if}$ is the dipole matrix element for the photo-excitation, $f(\omega)$ is the Fermi-Dirac function, and $A(\kvec, \omega) = \frac{1}{\pi} \Im G(\kvec, \omega)$ is the single particle spectral function, where $G$ is the single particle Green's function \footnote{We use the advanced Green's function as in Ref.~\onlinecite{gweon_extremely_2011}.}.

As the single particle Green's function in the normal state is believed to contain vital information on the nature of excitations relevant to the high temperature (``high \tc{}'') superconductivity, its characterization by ARPES has been a major line of research.  Various approaches towards getting at this information have been attempted: a phenomenological approach based on a simple scaling behavior of the electron self energy \cite{varma_phenomenology_1989}, an asymptotic solution to the Gutzwiller projected ground state of the \tj{} Hamiltonian \cite{anderson_hidden_2008}, application of a non-Fermi liquid theory \cite{orgad_evidence_2001} for low dimensions, and a newly proposed solution to the \tj{} Hamiltonian \cite{gweon_extremely_2011}.

For an experimental ``cut,'' i.e.\@ an experimental data set taken along a line of $\kvec$ values, $I(\kvec, \omega)$ is a function defined on a two dimensional domain.  This multi-dimensionality makes analyzing $I(\kvec, \omega)$ a non-trivial task.  While attempts \cite{meevasana_extracting_2008} have been made to analyze the $I(\kvec,\omega)$ image (e.g., see Fig.~\ref{fig-Bi2212data}(a)) as a whole, the current understanding of line shapes in terms of $A(\kvec, \omega)$ depends on the analysis of selected energy distribution curves (EDCs; EDC is a function of $\omega$, defined as $I(\kvec = \kvec_0,\omega)$) \cite{anderson_hidden_2008,gweon_extremely_2011,varma_phenomenology_1989,kaminski_momentum_2005} or selected momentum distribution curves (MDCs; MDC is a function of $\kvec$, defined as $I(\kvec,\omega = \omega_0)$, with $\kvec$ varying along a line) \cite{valla_evidence_1999,kaminski_momentum_2005}.

Currently, there is no consensus on a theoretical model that can suitably describe ARPES data of high \tc{} materials.  A model that can describe the normal state data, both EDCs and MDCs, obtained in different experimental conditions and for different materials, with the same intrinsic parameters would be a good candidate.  Here, we propose a new such phenomenological model.

The new model arises as the result of critically improving the so-called extremely correlated Fermi liquid (ECFL) model \cite{gweon_extremely_2011}, which was shown to be quite successful in describing EDCs.  The new model now makes it possible to describe other key aspects of the data as well: MDC fits are excellent and the values of $|M_{if}|^2$ behave reasonably.  And, it improves EDC fits, to boot.  {\em The result is a phenomenological model in which the apparent dichotomy between the EDCs and the MDCs} \cite{orgad_evidence_2001,gweon_generalized_2003} {\em are described excellently by two independent aspects of a single theoretical concept, the caparison factor} \cite{shastry_extremely_2011,gweon_extremely_2011}.

A phenomenological study of this kind seems to be helpful, also in light of the on-going development of the ECFL theory \cite{shastry_extremely_2013,hansen_extremely_2012}.  The theoretical formalism of ECFL initiated by Shastry \cite{shastry_extremely_2011,shastry_extremely_2013} is quite involved, and, while a numerical solution \cite{hansen_extremely_2012} valid for hole doping $x \gtrsim 0.3$ is now available, more time seems necessary to extend these promising results to near-optimal doping.  Thus, a phenomenological model based on the main feature of the theory, the caparison factor, may be of considerable value at this stage.  In this theory \cite{shastry_extremely_2013}, the caparison factor is an {\em $\omega$-dependent adaptive spectral weight} that encodes two key pieces of physics: the Gutzwiller projection that reduces the spectral weight at high $\omega$ and the invariance of the Fermi surface volume at low $\omega$.

In our previous work \cite{gweon_extremely_2011}, it has been demonstrated that the normal state EDCs for optimally doped cuprates for two different compounds, or for different experimental conditions (low photon energy or high photon energy), can be explained using an ECFL line shape model, all with one set of intrinsic parameters.  We will refer to that model as the ``simplified ECFL (\secfl)'' model \cite{shastry_shastry_2012}, in relation to the fuller theory in development \cite{shastry_extremely_2013,hansen_extremely_2012}.  While the EDC analysis used there has strong merits \cite{gweon_extremely_2011,casey_accurate_2008}, a natural subsequent question is whether MDCs can be described as well, along the same line of theory.

In the \secfl{} model \cite{gweon_extremely_2011}, $G(\kvec, \omega)$ is given by
\begin{align}
	G(\kvec, \omega) = \frac{Q_n - \frac{n^2}{4} \frac{\Phi(\omega)}{\Delta_0}}
		{\omega - \varepsilon(\kvec) - \Phi(\omega)}
\end{align}
where $ Q_n = 1 - \frac{n}{2} = \frac{1+x}{2} $ is the total spectral weight per $\kvec$ in the \tj{} model, and $n$ ($x$) is the number of electrons (holes) per unit cell \footnote{We now use the symbol $\varepsilon(\kvec)$, instead of $\xi(\kvec)$ (Ref.~\onlinecite{gweon_extremely_2011}), for the one electron energy.}.  $\Phi(\omega)$ is an ordinary Fermi liquid self energy, determined by two intrinsic parameters, $Z_{FL}$ (quasi-particle weight) and $\omega_0$ (cutoff energy scale), and one extrinsic parameter $\eta$ (impurity scattering contribution to $\Im \Phi$).  $\Delta_0$ is an energy scale parameter, determined completely by $n,Z_{FL}$, and $\omega_0$, through the global particle sum rule.  In Supplementary Materials (SM), a short summary of Ref.~\onlinecite{gweon_extremely_2011} is provided for readers' benefit.

\begin{figure}[t] 
\centerline{\includegraphics[width=3.3in]{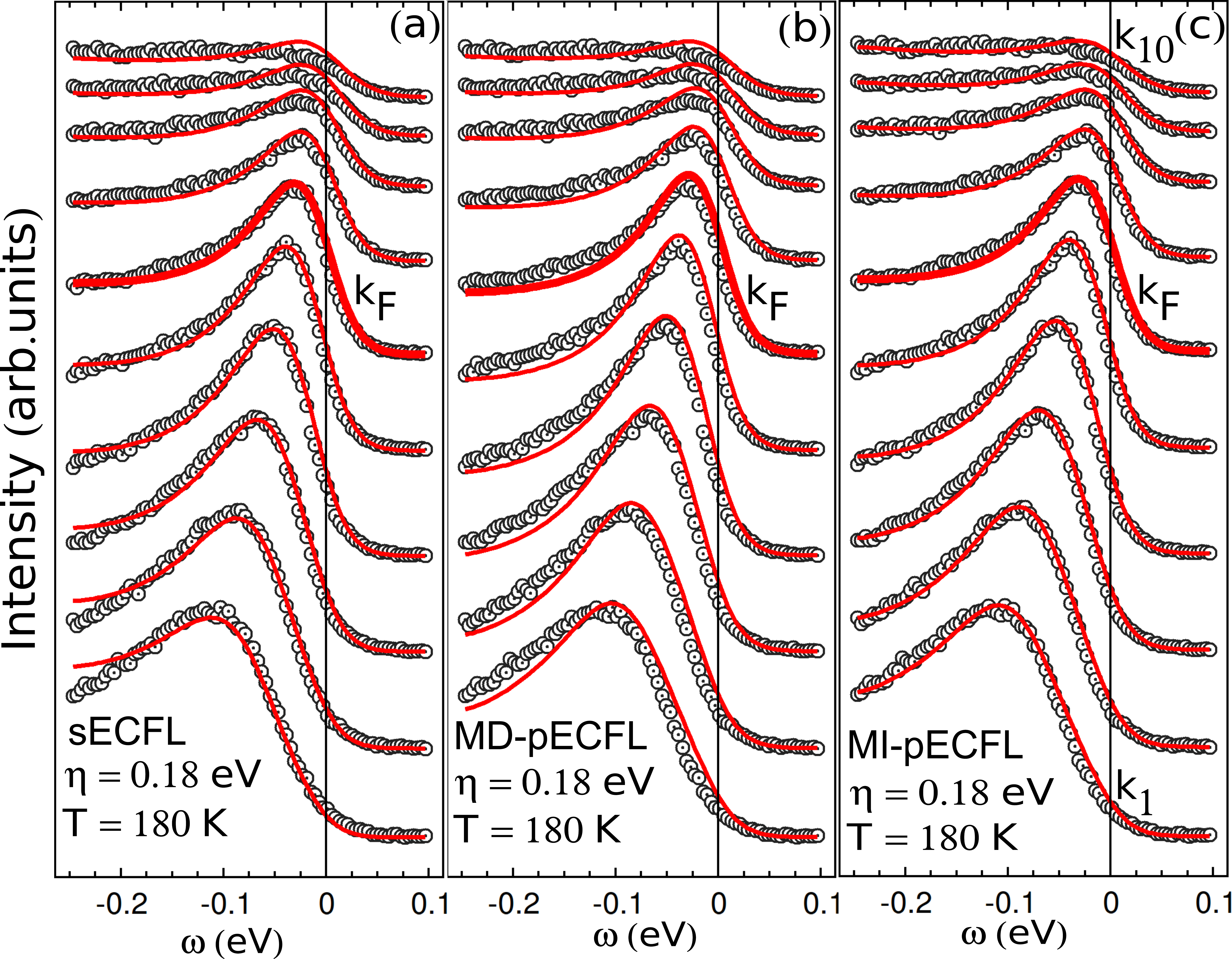}}
\caption{Line shape fits of EDCs for Bi2212 ($x =$ 0.15) using (a) \secfl{}, (b) \pecmd{}, and (c) \pecmi{}.  Data and model parameters are identical with those in Ref.~\onlinecite{gweon_extremely_2011} ($Z_{FL} = 0.33$, $\omega_0 = 0.5$ eV, $\Delta_0 = 0.12$ eV), except for slightly different values for $\eta$ ($0.17 \rightarrow 0.18$ eV) and $\varepsilon(\kvec)$ (see text).}
\label{fig-Bi2212EDC}
\end{figure} 

The above Green's function can be rewritten as
\begin{align}
	G (\kvec,\omega) &= \frac{Q_n}{\gamma_n} + \frac{\caparison}
		{\omega - \varepsilon(\kvec) - \Phi (\omega)}
		\label{eq-G} \\
	\caparison &= Q_n \left( 1 - \frac{\omega -
			\varepsilon(\kvec)}{\gamma_n} \right)
		\label{eq-caparison-factor}
\end{align}
where $\caparison$ is the ``caparison factor'' \cite{shastry_extremely_2011,gweon_extremely_2011} and the energy scale $\Delta_0$ is absorbed into $\gamma_n \equiv 4Q_n \Delta_0 / n^2$.  As all symbols in Eq.~\ref{eq-G} other than $\Phi(\omega)$ are real,
\begin{align}
	A(\kvec,\omega) &= \caparison\, A_{FL}(\kvec,\omega)
		\label{eq-A}
\end{align}
where $A_{FL}$ is the spectral function for the ``auxiliary Fermi liquid (AFL)'' Green's function \footnote{As in Ref.~\onlinecite{gweon_extremely_2011}, subscript ``FL'' means the AFL (auxiliary Fermi liquid), throughout this Letter.}, $A_{FL} = \frac{1}{\pi}\, \Im G_{FL} = \frac{1}{\pi}\, \Im [ \omega - \varepsilon(\kvec) - \Phi(\omega) ]^{-1}$.

The caparison function $\capa$, summarized concisely in Eq.~\ref{eq-caparison-factor}, played the central role in the sECFL model.  In this work, we show how its role can be extended even further by a key phenomenological modification: {\em inspired by data, we treat the $\omega$ dependence and the $\vec k$ dependence of $\capa$ as separately adjustable}.  We shall refer to the modified model as \pecfl{}, where p stands for ``phenomenological.''  We distinguish between \pecmd{} and \pecmi{} based on whether $\capa$ remains momentum dependent (MD) or made momentum-independent (MI).

With this much introduction to our models, we shall first discuss line shape fits, before discussing the models.  As for free fit parameters controlling the line shape, all models have $\eta$ and $\omega_0$, like \secfl{} \cite{gweon_extremely_2011} (cf.\@ SM Sec.~A).  In addition, the group velocity, $v_{F0}$, of $\varepsilon(\kvec)$, required small adjustment for different models to give correct peak positions (SM Sec.~B).  Then, only for \pecmd{}, there are two more free fit parameters (see later).

\begin{figure}[b] 
\centerline{\includegraphics[width=3.3in]{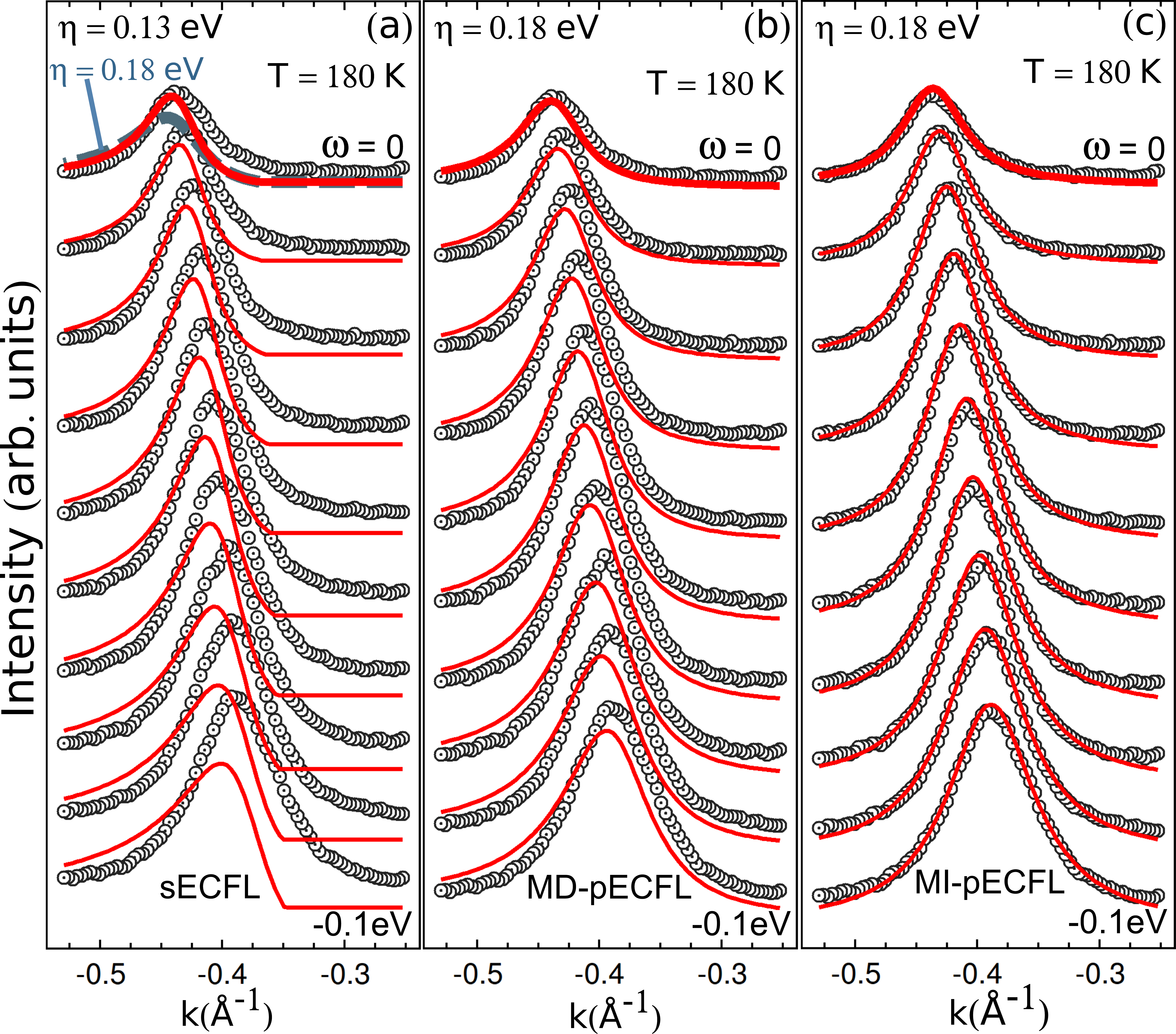}}
\caption{Line shape fits of MDCs for Bi2212 ($x =$ 0.15) using (a) \secfl{} and (b) \pecmd{} and (c) \pecmi{}.  Fit parameters are identical with those used for Fig.~\ref{fig-Bi2212EDC}, except for the reduced $\eta$ value (0.13 eV) for (a).}
\label{fig-Bi2212MDC}
\end{figure} 

Figure \ref{fig-Bi2212EDC} shows ARPES line shape fits for the normal state data for the optimally doped \bittot{} (Bi2212) sample, taken along the ``nodal direction,'' $(0,0)\rightarrow (\pi,\pi)$.  Panel a shows fits essentially identical \footnote{The slight difference is due to a slight change of $\eta$ (0.17 $\rightarrow$ 0.18 eV), noted in the caption of Fig.~\ref{fig-Bi2212EDC}.} with those of Ref.~\onlinecite{gweon_extremely_2011}.  The fit quality of \pecmi{} is clearly the best, while that of \pecmd{} is noticeably poorer, despite more fit parameters.

Figure \ref{fig-Bi2212MDC} shows ARPES line shape fits for MDCs of the same data set.  Panel a shows clearly that \secfl{} has difficulty fitting the data even at $\omega = 0$ (Fermi energy).  Panel b shows a quite improved fit by the \pecmd{} model.  However, the \pecmi{} fit shown in panel c is definitively the best.

\begin{figure}[b] 
\centerline{\includegraphics[width=3in]{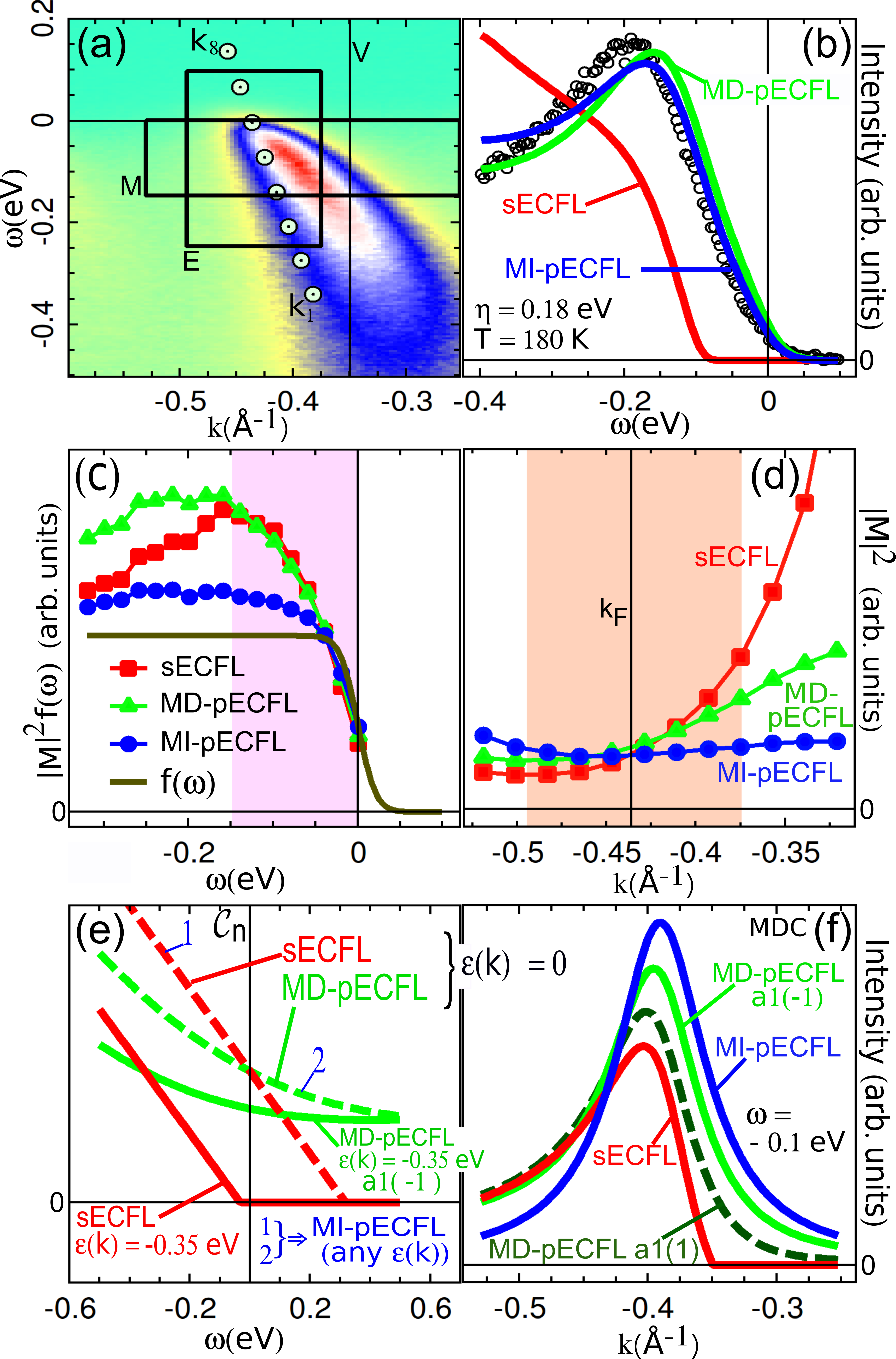}}
\caption{(a) The ARPES data for Bi2212, fit in previous figures.  Rectangle E (M) marks the range of data fit in Fig.~\ref{fig-Bi2212EDC} (\ref{fig-Bi2212MDC}).  Circle symbols mark $\varepsilon(\kvec)$ values used in the pECFL fits.  ARPES count increases from green, blue (half max), white, to red (max).
(b) EDC and its fits, for $\kvec$ value marked by the vertical line V in (a).
(c,d) The overall intensity scale parameters determined from the MDC fit (c) and the EDC fit (d), which correspond to $|M|^2f(\omega)$ and $|M|^2$, respectively, by Eq.~\ref{eq-sudden-approximation}.  Shaded areas marks the fit ranges used in Figs.~\ref{fig-Bi2212EDC} and \ref{fig-Bi2212MDC}.  As the energy dependence of $|M|^2$ is expected to be weak for this small range of $\omega$, we expect the points shown in panel c to approximately follow $f(\omega)$ (line).  \pecmi{} does this the best.  We also expect points in (d) to show only a modest variation in this $k$ range \cite{bansil_matrix_1998,meevasana_extracting_2008}.  Here also, \pecmi{} performs the best; in contrast, \secfl{} shows an unnatural steep increase.
(e) $\caparison$ for various models used. For \pecmd{}, $a_1 = -1$ and $a_2 = 2$ are used throughout this Letter.
  For \secfl{}, $\max{(\caparison,0)}$ is used \cite{gweon_extremely_2011}.
(f) The evolution of the MDC asymmetry, controlled by $a_1$ within \pecmd{} ($a_2 = 2$).  The MDC by \secfl{} is the most asymmetric, while that  by \pecmi{} is completely symmetric.
}
\label{fig-Bi2212data}
\end{figure}

That the \pecmi{} model is able to describe EDCs and MDCs so accurately seems to confirm the basic ECFL idea \cite{gweon_extremely_2011}.  In these fits, no extra component (e.g., extrinsic background intensity) was added to the theory that we described thus far \footnote{A small ``elastic background line shape'' (0.5 times the raw line shape for $k \gg k_F$) had been subtracted prior to fit, as explained in Ref.~\onlinecite{gweon_extremely_2011}.}.  All of the conclusions above also apply to the fits of the 91 K data \cite{gweon_extremely_2011}, as shown in detail in SM Sec.~C\@.

From the above work, it is clear that the \pecmi{} model emerges as the best model for the Bi2212 data.  This model is surprisingly simple: the $\varepsilon(\kvec)$ term in Eq.~\ref{eq-caparison-factor} is simply dropped.  The motivation for doing so is purely empirical: the MDCs of Bi2212 data are known to be quite symmetric and Lorentzian-like.  The effect of this simple modification is surprisingly very good in many ways.  MDC fits improve dramatically, as expected (Fig.~\ref{fig-Bi2212MDC}(c)), but EDC fits improve also (Fig.~\ref{fig-Bi2212EDC}(c)), especially for $\vec k$ far away from $\vec k_F$ (Fig.~\ref{fig-Bi2212data}(b)).  Furthermore, the overall scale parameters for MDC fits (Fig.~\ref{fig-Bi2212data}(c)) and EDC fits (Fig.~\ref{fig-Bi2212data}(d)) are now quite reasonable, as discussed in the caption.  These facts lend an overwhelming support to the \pecmi{} model.

The \pecmi{} model accomplishes these feats {\em without} any additional fit parameter, in comparison to the \secfl{} model.  Instead, the success arises crucially from {\em the separate treatment of the $\omega$ dependence and the $\kvec$ dependence, or independence, of the caparison factor, important for describing EDCs and MDCs, respectively}.

In contrast to the \pecfl{} models, it is clear that the \secfl{} model cannot describe MDCs at all.  Using identical fit parameters as for EDCs (see the dashed line marked ``$\eta = 0.18$ eV'' in Fig.~\ref{fig-Bi2212MDC}(a)), we get very poor fit quality, which improves, dramatically but insufficiently, by relaxing the $\eta$ parameter to 0.13 eV (Fig.~\ref{fig-Bi2212MDC}(a)).  In this new light, the \secfl{} model, so successful in the previous work \cite{gweon_extremely_2011}, must be viewed as getting only one of the two things correct---the $\omega$ dependence of the caparison factor, but not its $\kvec$ dependence---and its valid regime remains \cite{gweon_extremely_2011} confined to EDCs in the narrow range of $\kvec$ around $\vec k_F$ (Figs.~\ref{fig-Bi2212EDC}(a) and \ref{fig-Bi2212data}(a); see SM Sec.~E, also).

How about the \pecmd{} model?  From our discussion up to this point, it does not seem worth much consideration.  But, note that neither \secfl{} nor \pecmi{} guarantees the fundamental requirement $\caparison \ge 0$ (Fig.~\ref{fig-Bi2212data}(e)).  In the \pecmd{} model, we take
\begin{align}
\gamma_n &= \gamma_{n0} \left[ 1 + \exp
		\left( \frac{\omega - \varepsilon(\kvec)
			- a_1 \gamma_{n0}}{a_2 \gamma_{n0}} \right) \right]
\end{align}
where $\gamma_{n0} \equiv 4 Q_n \Delta_0 / n^2 = 0.38$ eV is the value of $\gamma_n$ in the \secfl{} model.  In \pecmd{}, $\caparison \ge 0$ {\em is} guaranteed for any $\kvec$ and $\omega$ values, if $a_1 \le 1 + a_2\, (1 - \log a_2)$.  Physically, $a_1$ and $a_2$ play the role of controlling the MDC asymmetry (Fig.~\ref{fig-Bi2212data}(f)) and were determined as $a_2 = 2 \pm 1$ and $a_1 = -1 \pm 1$.  See SM Sec.~F for details.

Accordingly, $\capa$ for \pecmd{} stays clearly above zero and is smooth (Fig.~\ref{fig-Bi2212data}(e)).  $\capa$ for \pecmi{} is, by definition, that for \secfl{} at $\ek = 0$, as marked by label 1 in Fig.~3(e).  However, we find that it can also be taken to be that for \pecmd{} at $\ek = 0$, as indicated by label 2 in Fig.~\ref{fig-Bi2212data}(e), since fit results are very comparable between these two choices.

For Bi2212, \pecmd{} is significantly better than \secfl{}, but significantly worse than \pecmi{}, despite having two more fit parameters (cf.\@ SM Sec.~F).

\begin{figure}[t] 
\centerline{\includegraphics[width=3in]{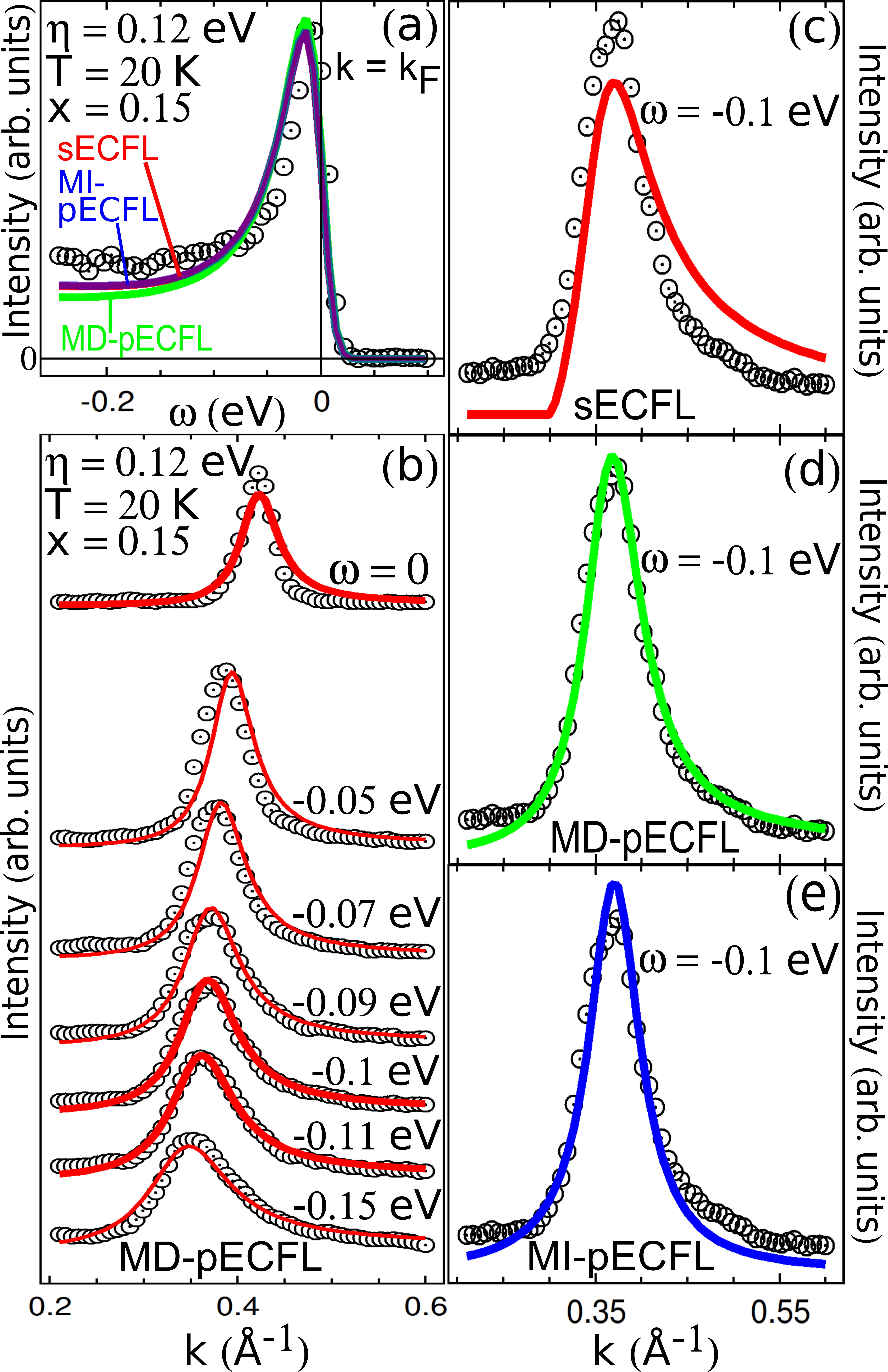}}
\caption{Fits to the data of optimally doped ($n = 0.85$; $x = 0.15$) LSCO \cite{yoshida_low-energy_2007}, taken along the nodal direction.  (a) EDC fits at $k=\kf$.  (b,c,d,e) MDCs for $-\omega \gtrsim 0.07$ eV are significanly asymmetric, described the best by \pecmd{}.}
\label{fig-LSCO}
\end{figure}

However, the situation changes when we fit data of \lsco{} (LSCO) \cite{yoshida_low-energy_2007}, showing strong MDC asymmetry (panels b--e).  Here, identical fit parameter values as those for Bi2212 are used, except for $\eta = 0.12$ eV and $v_{F0}$ (SM Sec.~B).  Fig.~\ref{fig-LSCO}(a) shows an EDC fit, good by all models, just as for Bi2212.  However, the MDC fit is a different matter.  Notably, MDCs show significant asymmetry for $-\omega \gtrsim 0.07$ eV (panel b), and that asymmetry can be described properly only by the \pecmd{} model, as illustrated clearly in fits shown in panels b through e, and as discussed further in Sec.~D of SM\@.

We see that the original \secfl{} model must be modified greatly (Bi2212) or somewhat (LSCO) to describe MDCs.  We argue that these phenomenological modifications require physics beyond the \tj{} Hamiltonian, since the \secfl{} model is derived \cite{shastry_extremely_2011} from the \tj{} Hamiltonian, and another well-known model \cite{casey_accurate_2008} based on the \tj{} Hamiltonian also implies too asymmetric MDCs.  More specifically, the physics of the (next nearest neighbor hopping) $t^\prime$ term seems a good candidate: the well-known fact that $|t^\prime/ t|$ is significantly smaller for LSCO \cite{markiewicz_one-band_2005,pavarini_band-structure_2001} goes well with our result that the \pecmd{} model is more similar to the \secfl{} model.  The $t^\prime$ term is correlated with the superconducting transition temperature \cite{pavarini_band-structure_2001}, imparting importance to our current proposal.  We recently found \cite{gweon_spectroscopic_2013} that an anomalous ARPES feature is explained by pECFL, but not \secfl{}, and has similarity to a scanning tunneling spectroscopy feature correlated with superconductivity, adding more credence to our argument here.  Lastly, the fact that the caparison factor for the infinite-dimensional ECFL becomes $\kvec$-independent \cite{zitko_extremely_2013,perepelitsky_extremely_2013} seems to go along with our result, within the crude analogy between adding a large $t^\prime$ term and increasing channels for $t$ hopping.

In this Letter, we proposed a phenomenological ARPES line shape model, based on the ECFL theory \cite{shastry_extremely_2011,shastry_extremely_2013}.  The essential feature of our model remains the caparison factor \cite{shastry_extremely_2011,gweon_extremely_2011,hansen_extremely_2012}, which is capable of describing both anomalous EDC line shapes \cite{casey_accurate_2008,gweon_extremely_2011}, universal for \hightc{} cuprates, and apparently more conventional MDC line shapes \cite{kaminski_momentum_2005,valla_evidence_1999}.  While our model is not the first to fit both EDCs and MDCs \cite{kaminski_momentum_2005} of \hightc{} cuprates, its demonstrated fidelity (including a qualitative description of $|M_{if}|^2$) and range of applicability is now unprecedented.  Also unprecedented is the notable fact that our model requires a Dyson self energy \cite{shastry_anatomy_2011}, whose form is {\em drastically different from that assumed by the prevalent, but incomplete, MDC-only analysis} \cite{kokalj_consistent_2011,chang_anisotropic_2008}: to our knowledge, ours is the only $\vec k$-dependent \cite{shastry_anatomy_2011} Dyson self energy that has fit cuprate MDCs.  Thus, extending the current analysis to wider ranges of momentum, doping, and temperature and studying its implication on other properties such as the resistivity \cite{hansen_extremely_2012} seems to make a great research topic for the immediate future.

\begin {acknowledgments} 

We gratefully acknowledge B. S. Shastry and D. Hansen for stimulating discussions and feedback to the manuscript, and G. D. Gu for Bi2212 samples used for the original data \cite{gweon_extremely_2011}.  We thank T. Yoshida for sharing the digital version of the LSCO data.  The work by GHG was supported partially by COR-FRG at UC Santa Cruz.  Portions of this research were carried out at the SSRL, a Directorate of SLAC National Accelerator Laboratory and an Office of Science User Facility operated for the U.S. DOE Office of Science by Stanford University.

\end {acknowledgments} 

\clearpage

\renewcommand{\thefigure}{S\arabic{figure}}
\setcounter{figure}{0}

\centerline{\Large \bf Supplementary Materials}
\vspace{2em}

One of the main references in this work is the previous work by Gweon, Shastry and Gu (GSG), reference 7 of the main text.  In this document, we shall refer to this reference as GSG\@.

Before we present our materials, let us note that this document introduces new figures. 
For figure labels, we use prefix ``S'' followed by the figure number for figures introduced in this document, e.g., Fig.~S1, to make them clearly distinguishable from the figure labels of the main text, e.g., Fig.~1.  This document does not introduce any new equation numbers or reference numbers.

\section{Summary of key results of GSG} 

It is for the probable benefit to readers that we now summarize key results of the previous work in GSG\@.  The Dyson self energy of the AFL, $\Phi$, is given by (as given by Eq.~2 and footnote 6 of GSG)
\begin{align*}
	\Im \Phi (\omega) &= \frac{\omega^2 + \tau^2}{\Omega_0}\, e^{-\frac{\omega^2 + \tau^2}{\omega_0^2}} + \eta, \\
	\Re \Phi (\omega) &= \frac{-1}{\sqrt{\pi}\, \Omega_0}\, e^{-\frac{\tau^2}{\omega_0^2}}\, \left[ \omega_0 \omega - 2 (\omega^2 + \tau^2) D \left( \frac{\omega}{\omega_0} \right) \right],
\end{align*}
where $\tau = \pi k_B T$, $Z_{FL} = \left( 1 + \frac{\omega_0}{\sqrt {\pi} \Omega_0} \right)^{-1}$, and $D(x) = \frac{\sqrt{\pi}}{2} e^{-x^2} \text{erfi} (x)$ is the Dawson function.  EDC fits in GSG used fixed values of $Z_{FL} = 1/3$ and $\varepsilon (\kvec)$.  The experimental data that helped fix these values are the ARPES data taken up to very high binding energy (Ref.~8).  Using such data it was possible to find the quasi-particle dispersion renormalization ratio due to the high energy ARPES ``kink,'' $Z_{FL}$.  It was also possible to fix $\varepsilon (\kvec)$, by taking it to have the same form as the tight-binding band dispersion well-known in the literature, which was scaled to give the correct bandwidth measured by the ARPES data.  This left only two free fit parameters, $\eta $ and $\omega _0$. Of these two, $\eta $ is the impurity scattering parameter, which we associated with the effective sample quality probed by the ARPES technique under different conditions; it is determined in practice by the width of the sharpest quasi-particle peak of a given data set, $\approx 0.04$ eV for data taken with low energy photons (e.g., from laser) and 3 or 4 times greater for data taken with high energy photons at synchrotrons.  Thus, the only intrinsic fit parameter was $\omega_0$, which was determined from fits as $\omega _0 = 0.5 \pm 0.1$ eV\@.  Then, through a theoretical constraint equation, $\Delta _0 = 0.12 \pm 0.02$ eV  was determined. GSG were able to successfully interpret $\omega _0$ and $\Delta _0$ as the purely electronic high and low ARPES kink energy scales, respectively (Fig.~5 of GSG).  This interpretation provided these two energy scale parameters with their physical meanings.  In the current work, we use the values of all of the intrinsic parameters ($Z_{FL}, \omega _0, \Delta _0$) without change from GSG, as we found that all fits were stable under small variations of these parameters.  However, small variations of $\varepsilon(\kvec)$ were necessary, as described in the next section.

\section{One electron dispersion relation} 

The key information on the one electron dispersion relation, $\varepsilon (\kvec)$, is provided in the main text as circle symbols in Fig.~3(a).  Here, we provide more detailed information.  For all fits reported in this work, $\varepsilon (\kvec)$ could be approximated as a line, $\varepsilon (\kvec) = v_{F0} (k-k_F)$, due to the small energy range and the momentum range involved, as far as $\varepsilon (\kvec)$ is concerned.  We found that the model-dependent variation on $v_{F0}$ was required in order for the fits to describe experimental peak positions correctly. For fitting the Bi2212 data, $v_{F0} =$ 5.5 (\secfl{}) and 6.3 eV\r A (\pecmd{}, \pecmi{})\@.  In Fig.~3(a), $\varepsilon(\kvec)$ values for \pecfl{} fits are shown.  For fitting the LSCO data in Fig.~4, $v_{F0} =$ 4 (\secfl{}), 5 (\pecmd{}), and 5.5 eV\r A (\pecmi{}) were used.  This variation in $v_{F0}$ is consistent with the amount of uncertainty in knowing the precise band width (Ref.~8); it implies small additional uncertainties for the intrinsic parameters $Z_{FL}$, $\omega_0$, and $\Delta_0$, relative to those discussed in the previous section.

\section{Fits of the Bi2212 91K data} 

In the main text, we focused on the 180 K data of Bi2212.  Here, we show that the ARPES data at 91 K, right at $T_c$, of this material, is described equally well by the \pecmi{} model.

\begin{figure}[t] 
\centerline{\includegraphics[width=3.3in]{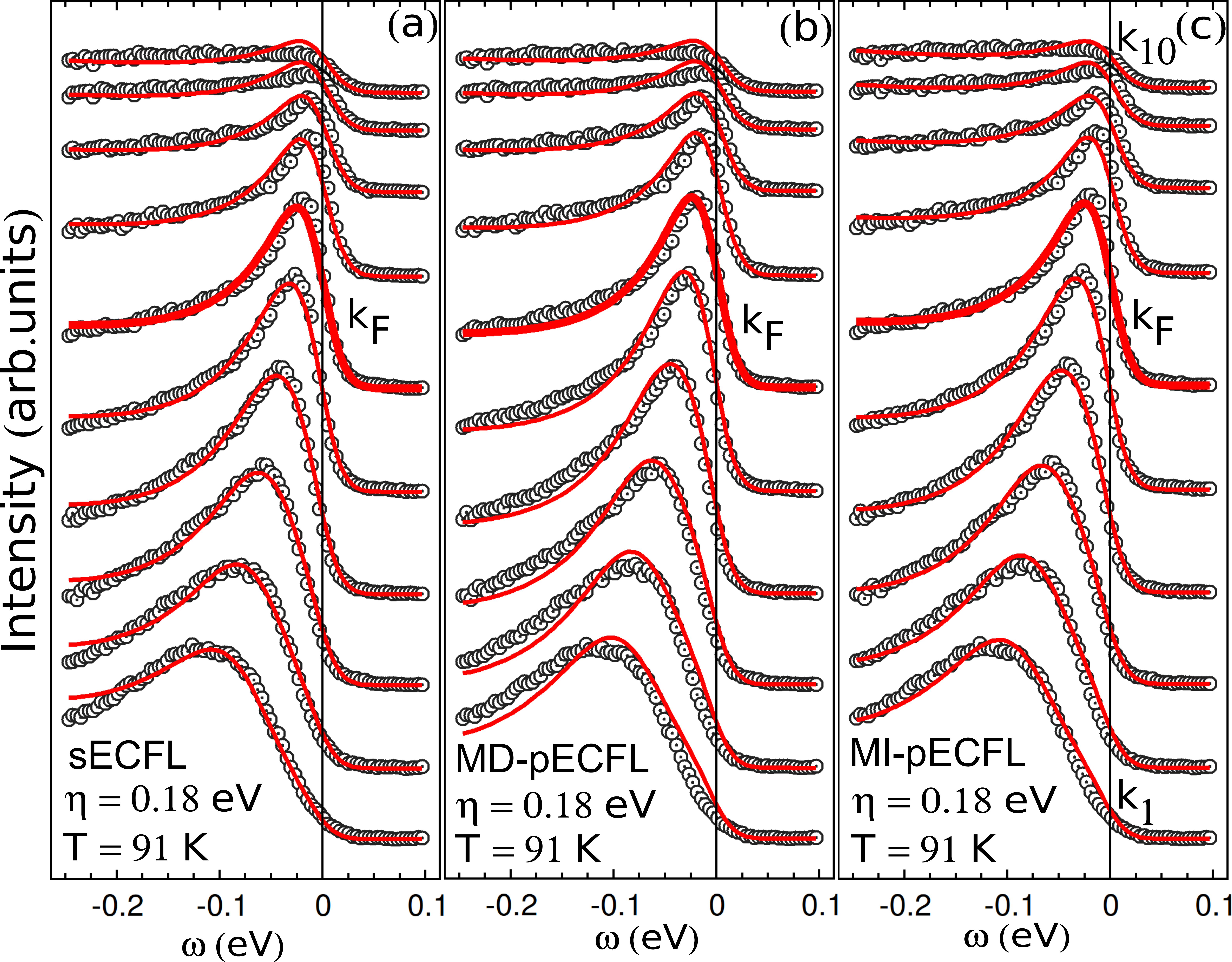}} 
\caption{Line shape fits of EDCs for Bi2212 ($x =$ 0.15) 91 K using (a) \secfl{}, (b) \pecmd{}, and (c) \pecmi{}.  Except the temperature and the overall intensity scale (see Fig.~\ref{fig-Bi2212-91K-weights}), all parameter values used for the fit are the same as those used for the 180 K data (Figs.\@ 1--3).}
\label{fig-Bi2212EDC-91K}
\end{figure}

\begin{figure}[b] 
\centerline{\includegraphics[width=3.3in]{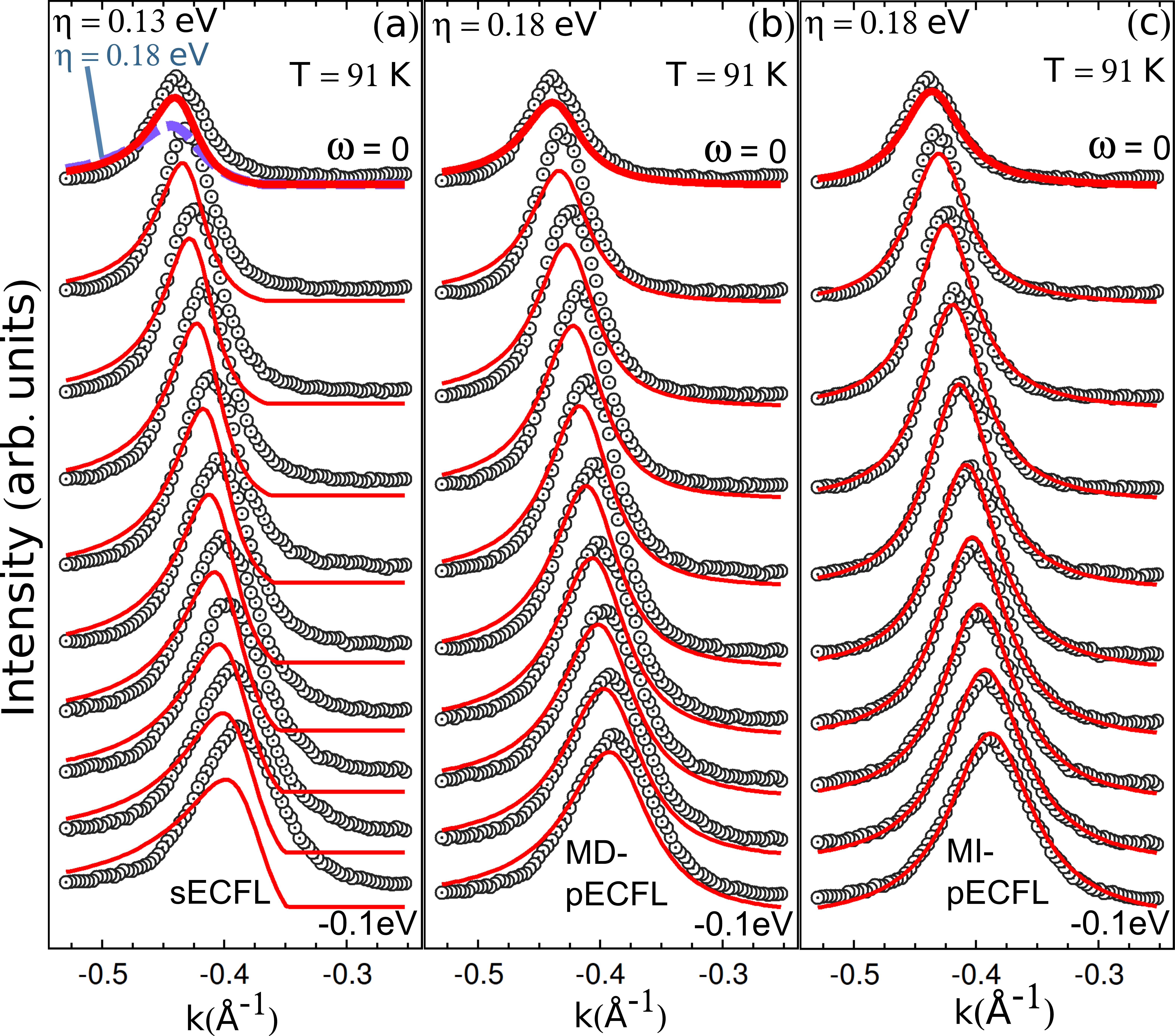}}
\caption{Line shape fits of MDCs for Bi2212 ($x =$ 0.15) measured at 91 K, using (a) \secfl{} and (b) \pecmd{} and (c) \pecmi{}.  Except the temperature and the overall intensity scale (see Fig.~\ref{fig-Bi2212-91K-weights}), all parameter values used for the fit are the same as those used for the 180 K data (Figs.\@ 1--3).}
\label{fig-Bi2212MDC-91K}
\end{figure}

Fig.~\ref{fig-Bi2212EDC-91K} shows that the \pecmi{} model describes the EDCs the best, while the other two models are still quite good.  Fig.~\ref{fig-Bi2212MDC-91K} shows that the \pecmi{} model describes the MDCs the best, while the other two models come significantly short.  In particular, the \secfl{} model does very poorly, even when the $\eta$ parameter is reduced, against the principle that both EDCs and MDCs must be described by the same set of parameters (cf.\@ Fig.~\ref{fig-Bi2212EDC-91K}), just to help the fit, as shown in panel a.

\begin{figure}[t] 
\centerline{\includegraphics[width=\columnwidth]{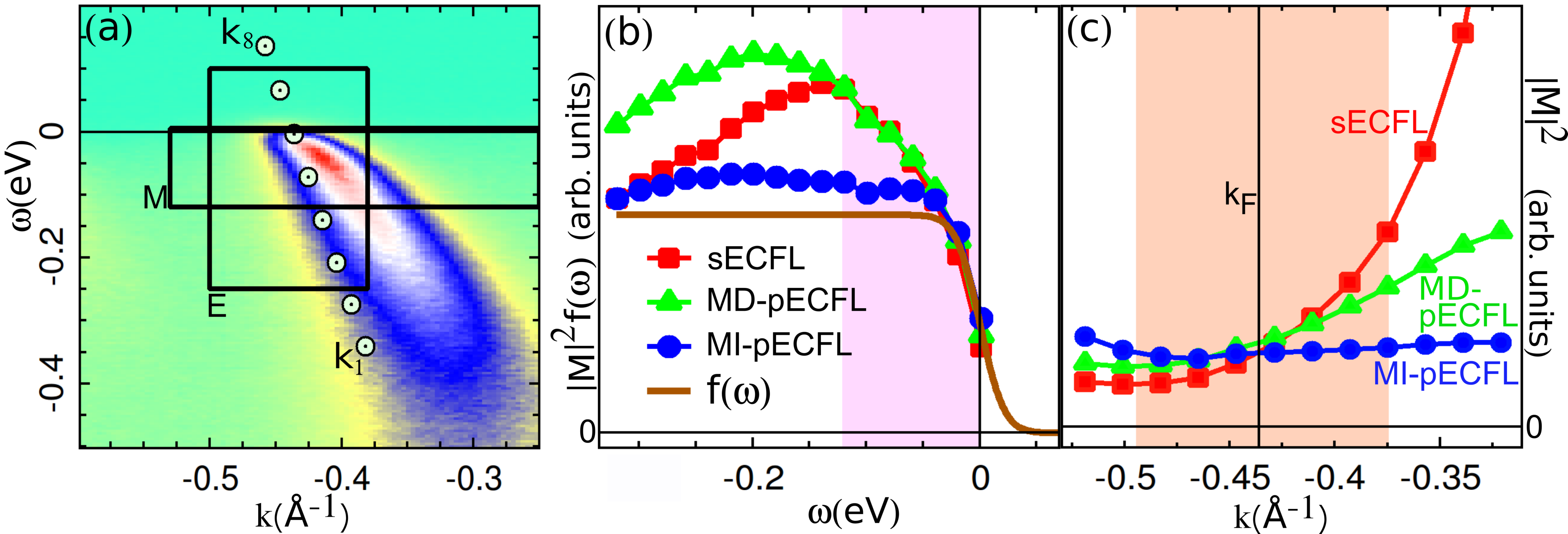}}
\caption{The ARPES data for Bi2212 at 91 K, fit in previous figures, represented in a false color scale (panel a) and the overall intensity scales used for the MDC fits (panel b) and the EDC fits (panel c).  Panels a, b, and c are presented in the same format as panels a, c, and d, respectively, of Fig.~3.  As in that figure of the main text, we see that the \pecmi{} model gives the most reasonable values of the overall intensity scales for the MDC fits (panel b) and the EDC fits (panel c), with only a mild dependence of the matrix element on the momentum and the energy, as anticipated.}
\label{fig-Bi2212-91K-weights}
\end{figure}

Fig.~\ref{fig-Bi2212-91K-weights} shows the 91 K data as an image, and also shows the overall scale factors, $|M|^2 f(\omega)$ as a function of $\omega$ and $|M|^2$ as a function of $\kvec$, as extracted from MDC fits and EDC fits, respectively.  In great agreement with Fig.~3, we see that the \pecmi{} model implies the most reasonable trend for the matrix element.

All of these findings are in complete agreement with those findings based on the 180 K data, discussed in the main text.

Note, however, that 91 K is the transition temperature of this material.  As the ECFL theory is the theory of the normal state, there is a reason to doubt or question the applicability of the model at very low energies at this temperature.  For instance, the rather noticeable discrepancy between the theory MDC and the measured MDC at $\omega = 0$, as can be noticed in Fig.~\ref{fig-Bi2212MDC-91K}(c), may be related to the inadequate nature of the theory at low energies for temperatures close to, or below, the superconducting transition temperature.  We discuss related issues in more detail in the next section.

\section{Low temperature data} 

\begin{figure}[t] 
\centerline{\includegraphics[width=3.5in]{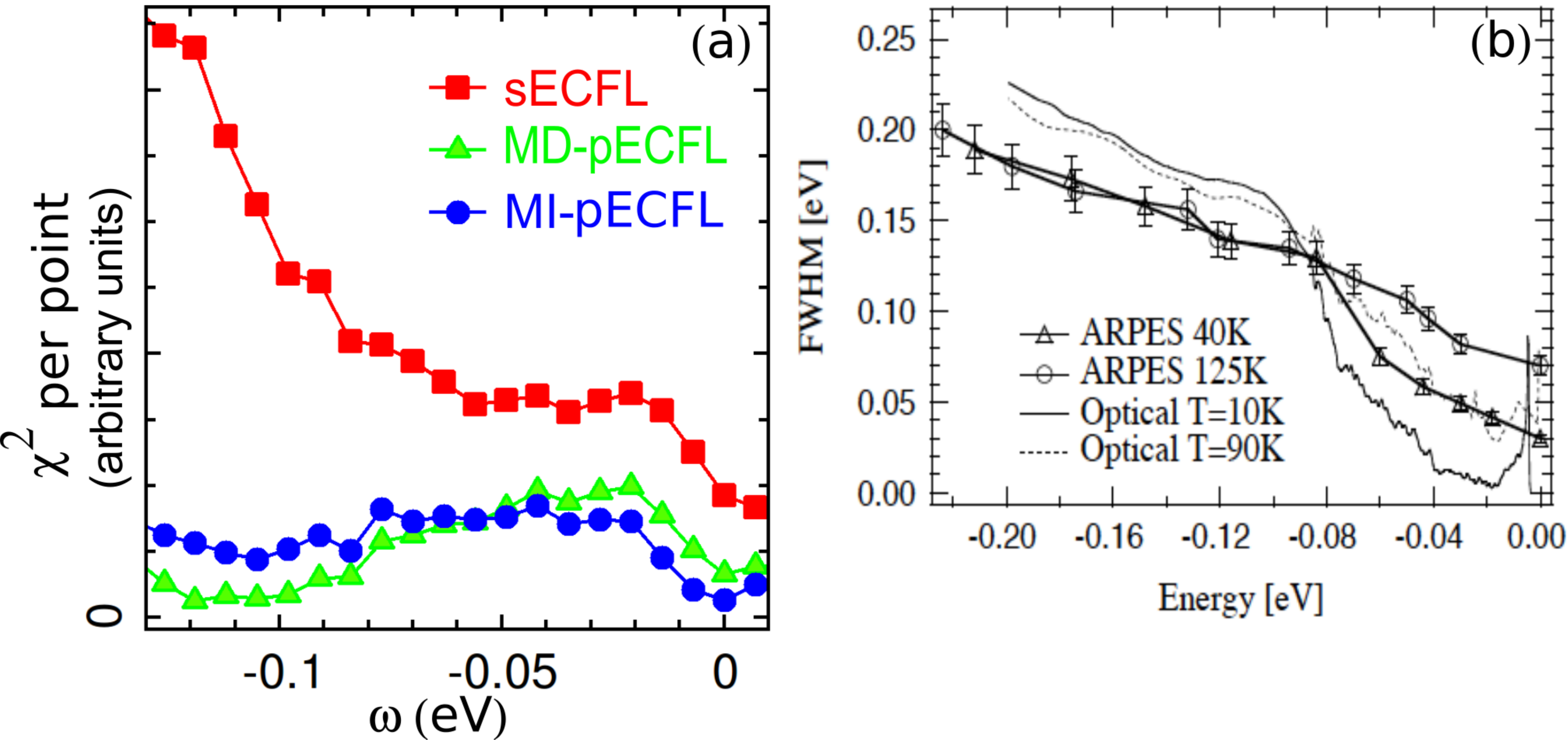}}
\caption{(a) $\chi ^2$ per point for the three models used for the line shape analysis of LSCO (25K) in Fig.~4.  (b) The FWHM values of MDCs for an optimally doped Bi2212 ($T_c = 90$ K), as reported by Kaminski et al., in Fig.~4 of their paper, Phys.~Rev.~Lett.~{\bf 84}, 1788 (2000).  Also included are the carrier scattering rate from infrared reflectivity measurements (A. V. Puchkov, et al., J. Phys.~Cond.~Matt.~{\bf 8}, 10049 (1996)).}
\label{FigF}
\end{figure}

In Fig.~\ref{FigF}(a), we report the quality of the MDC fits shown in Fig.~4, using the standard $\chi^2$ measure of the line shape fit.  Perfectly consistent with Figs.~4(c--e), we see that the $\chi^2$ values for the \secfl{} model is much too high compared to the $\chi^2$ values for the two \pecfl{} models.  This lends a definite support to the \pecfl{} model.  However, which \pecfl{} model is better?

As Fig.~\ref{FigF}(a) shows, the $\chi^2$ value is smaller for the \pecmi{} model at low energy ($-\omega \lesssim 0.05$ eV), while it is smaller for the \pecmd{} model for ($-\omega \gtrsim 0.05$ eV)\@.  So, our question above can be rephrased as ``which of these two facts is more significant?''  The answer is the latter, since the theory that we are applying here is applicable only to physics in the high temperature scale or the high energy ($-\omega$) scale.  When we examine the low temperature data, and compare the data with the theory for the normal state at high temperature, our primary concern must be focused on the high energy ($-\omega$) behavior.

This view point is also supported well empirically by the known results of the literature.  For instance, in Fig.~\ref{FigF}(b), we show a plot taken from a reference, given in the caption.  This plot shows that the FWHM values of the ARPES data are nearly temperature independent at high energy, $-\omega \gtrsim 0.08$ eV\@.  This is due to the fact that the high energy data taken at low temperatures are essentially identical with the high energy data taken at high temperatures.

This fact justifies our use of the low temperature data in Fig.~4.  It also guides us importantly to assess the relative merits of the two models, \pecmd{} and \pecmi{}, for the LSCO data.  Focusing on high energy, we conclude that the \pecmd{} model is much better, as its $\chi^2$ value is several times smaller than that for \pecmi{} at high energy $-\omega \gtrsim 0.05$ eV (Fig.~\ref{FigF}(a)).

This does not mean that we do not need to study the high temperature data for LSCO\@.  On the contrary, the current lack of the high temperature data for LSCO in view of our current result makes the study of high temperature ARPES data for LSCO a very attractive topic.  In view of the success of the ECFL model, the strong temperature dependence predicted at high temperatures (Fig.~5(f) of GSG), and the apparent difference between the LSCO data and the Bi2212 data, the investigation of the high temperature ARPES line shapes for LSCO may be an excellent topic of research in the near future.

Lastly, note that at low temperatures, the low energy ARPES peak is significantly sharper than our theoretical model predicts, as the MDC comparison at $\omega = 0$ in Fig.~4(b) or Fig.~\ref{fig-Bi2212MDC-91K}(c) shows.  This seems consistent with the general notion that the onset of the superconducting gap (or pseudo-gap) reduces the decay channels for quasi-particles.

\section{\secfl{} from \pecfl{}} 

Both the \pecmd{} model and the \pecmi{} model agree with the \secfl{} model, when small energy range and small momentum range are considered, as the EDC fits of Figs.~1 and \ref{fig-Bi2212EDC-91K} show.  A more quantitative analysis can be made about this fact, as follows.

As long as the caparison factor defined in Eq.~4 is well above zero, the \secfl{} model remains valid.  Using $\gamma _{n} = 4 Q_n \Delta_0 / n^2 = 0.38$ eV, determined from our fit parameters, and putting $\omega \approx Z \varepsilon (\kvec)$, the approximate quasi-particle peak position, we see that Eq.~4 requires $(Z - 1) \varepsilon (\kvec) \lesssim \gamma_n$ in order for $\caparison > 0$.  Here, we must take $Z$ to be the {\em total} mass renormalization ratio, which, according to our theory, is equal to $Z = Z_{FL} Q_n = Q_n/3 = 0.19$.  Putting these numbers together, we see that the \secfl{} model is expected to be valid for $\varepsilon (\vec{k}) \gtrsim -0.47$ eV, or $Z \varepsilon (\vec{k}) \gtrsim -0.09$ eV.  Noting that $Z\varepsilon (\kvec)$ corresponds to the approximate peak position in EDC, we therefore learn that the \secfl{} model would be good for describing EDCs, when their peaks are within about 90 meV from the chemical potential.  This explains why \secfl{} fits are very good in Figs.~1 and \ref{fig-Bi2212EDC-91K}, while this is no longer the case in Fig.~3(b).  It is for the same reason that MDC fits by the sECFL are very poor.

This consideration explains excellent EDC fits in GSG, justifies the simple truncation procedure, $\mathcal {C}_n\rightarrow \qopname \relax m{max}(\protect \mathcal {C}_n, 0)$, employed in that work (see Fig.~3(e) also), and, last but not least, correctly puts \pecfl{} models as standing in harmony with the \secfl{} model, not in contradiction to it.

\section{\pecmd{} and MDC asymmetry} 

The \pecmd{} model is useful to discuss from three points of view.  First, its discussion touches upon some basic theoretical issues.  Second, it provides an alternative way to define the \pecmi{} model (see Fig.~3(e) and its discussion in the main text).  Third, it provides the best model for the LSCO data as shown in Fig.~4.

Here, our discussion is mainly on the first point, since the other two points have been discussed already in the main text.

From Eq.~5, the non-negativity of the spectral function requires that $\caparison \ge 0$, or equivalently, $\gamma_n \ge \omega - \ek$.  Considering $\ek < 0$ and $\omega$ near the peak position ($\omega \approx Z\ek$ where $Z$ is the quasi-particle weight), this leads to the following requirement: $\gamma_n \gtrsim (1-Z)|\ek|$.  This requirement would clearly be violated for a large value of $|\ek|$, if $\gamma_n$ were to be held constant.  Going beyond the \secfl{} approach, where $|\varepsilon(\kvec)|$ was limited to a small value so that this violation is irrelevant (Sec.~E), we can avoid this violation altogether, if we employ a $\kvec$ or $\omega$ dependent $\gamma_n$.  In particular, in \pecmd{}, we take
\begin{align*}
\gamma_n &= \gamma_{n0} \left[ 1 + \exp
		\left( \frac{\omega - \varepsilon(\kvec)
			- a_1 \gamma_{n0}}{a_2 \gamma_{n0}} \right) \right].
\end{align*}
This is Eq.~6 of the main text.

Here, $a_2 > 0$ defines the width ($a_2 \gamma_{n0}$) of the sigmoidal drop of $1/\gamma_n$, occurring at a position defined by $a_1$: $\omega = \ek + a_1 \gamma_{n0}$.  This $\gamma_n$ function ensures that (1) $\capa \rightarrow Q_n/\omega$ as $\omega \rightarrow \infty$, as well as $\omega \rightarrow -\infty$, as required by the spectral weight sum rule per $\kvec$ within the \tj{} model, and (2) $A(\kvec, \omega) \ge 0$ for {\em any} $\kvec, \omega$ values, as long as
\begin{align*}
	a_1  \le 1 + a_2\, (1 - \log a_2) \, & \equiv\, a_{1,max}(a_2),
\end{align*}
as can be shown by a bit of algebra.

{\em This way, \pecmd{} ensures the sum rule and the non-negativity of $A(\kvec, \omega)$}.

The new parameters $a_1$ and $a_2$ play the role of controlling the MDC asymmetry.  Before we discuss this, we must first note that, for a given value of $a_2$, $a_1$ has both an upper bound, as just given, and a rough lower bound, $a_1 \gtrsim -a_2$.  The lower bound arises due to the empirical fact that the EDC line shape cannot be fit within the AFL theory, which the \pecmd{} theory converges to (up to an overall scale) if $a_1 \rightarrow -\infty$.  That is, $a_1$ cannot be too small.  These bounds and the line shape fit severely restrict values of $a_1$ and $a_2$: $a_1$ lies at about -1 and $a_2$ lies at about 2, both with a small wiggle room of about $\pm 1$.

The role of $a_1$ has been illustrated in Fig.~3(f).  As $a_1$ is tuned from $-\infty$ to $\infty$, the MDC goes from completely symmetric, just as in \pecmi{}, to very asymmetric, just as in \secfl{}.

\begin{figure}[t] 
\centerline{\includegraphics[width=3.3in]{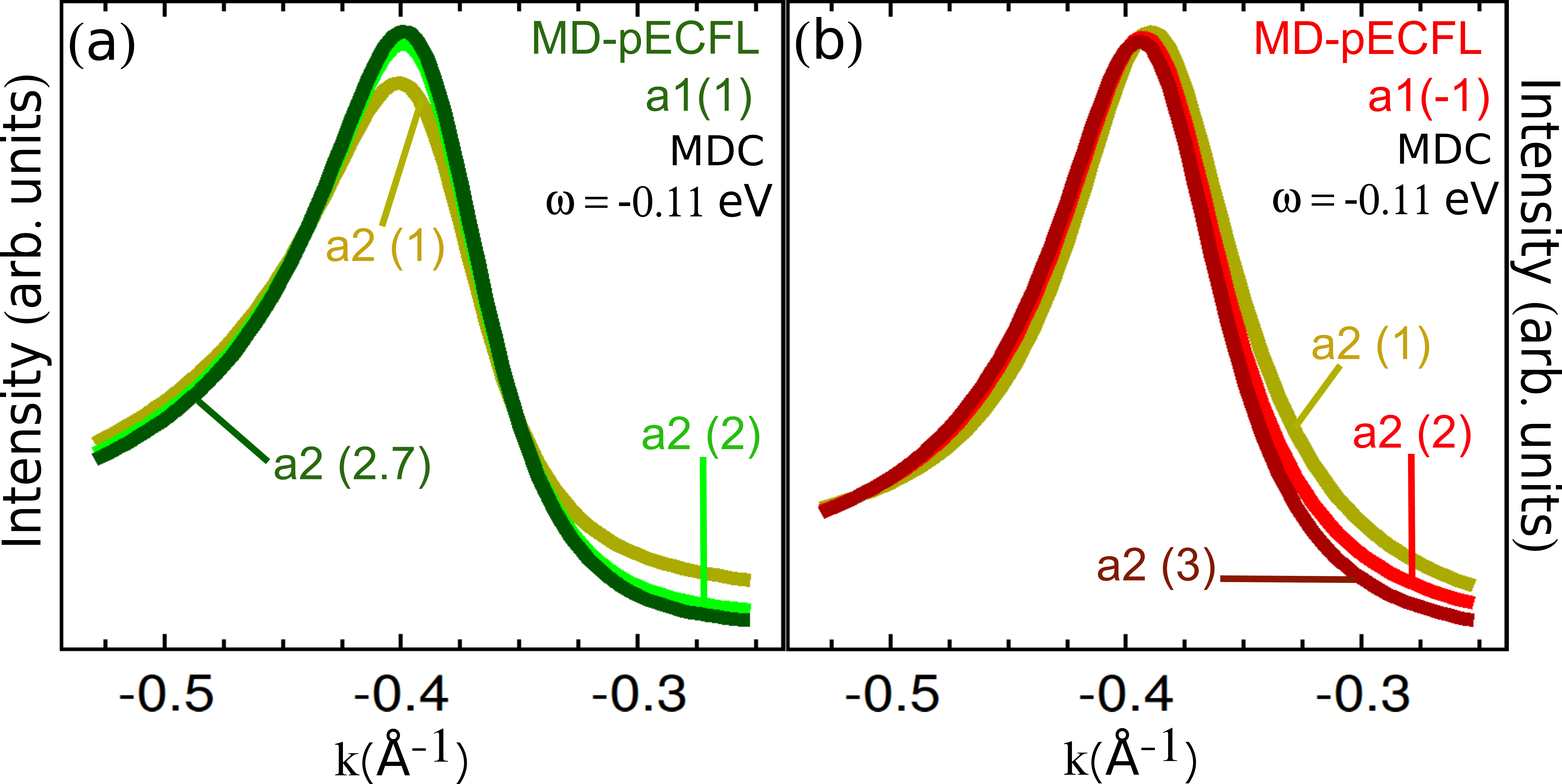}}
\caption{The MDC line shape dependence of \pecmd{} on the parameter $a_2$ at $\omega = -0.11$ eV,  (a) $a_1 = 1$, (b) $a_1=-1$.}
\label{fig-MDC-a2}
\end{figure} 

In this consideration, the following points are worth noting.  First, if $a_1 \rightarrow \infty$, then \pecmd{} converges to \secfl{}.  This is because $\gamma_n \rightarrow \gamma_{n0}$ in this limit (Eq.~6).  Second, if $a_1 \rightarrow -\infty$, then \pecmd{} does {\em not} convert to \pecmi{}, but to a ``re-weighted'' AFL.  This is because in this case $\caparison \rightarrow Q_n$ and $\gamma_n \rightarrow \infty$, giving $G(\kvec, \omega) = Q_n G_{FL} (\kvec, \omega)$, according to Eqs.~3 and 4, where $G_{FL}$ is the Green's function for the AFL, as defined right after Eq.~5.  So, the theory becomes that of the AFL theory, in this limit, except that the total spectral weight is not 1, but the \tj{} model weight $Q_n$---so, a ``re-weighted'' AFL\@.  Now, the \pecmi{} model and the AFL model are quite different, since the caparison factor gives a stronger EDC asymmetry in the former model. However, the two models do have in common that their MDCs are completely symmetric.  So, when narrowly focusing on the MDC asymmetry, we can view the three models---the \pecmi{} model, the AFL model, and the $a_1 = -\infty$ \pecmd{} model---as mutually equivalent.

From this discussion, the following can be noted for the reason why the Bi2212 data are fit poorly by the \pecmd{} model, despite more parameters afforded by it.  The combination of the asymmetric EDC behavior and the symmetric MDC behavior displayed by the Bi2212 data pose difficulties to the \pecmd{} model.  This is because as the MDC is made more symmetric, the EDC is becoming more symmetric also, contributing the degradation of the fit (Fig.~1(b) and Fig.~\ref{fig-Bi2212EDC-91K}(b)).  Separately, the $a_1 \rightarrow -\infty$ \pecmd{} model gives as good an MDC fit as the \pecmi{} model, while the $a_1 = 1$ \pecmd{} model gives as good an EDC fit as the \pecmi{} model.  The middle ground is found at $a_1 = -1$, at which value both the EDC fit and the MDC fit suffer a little.

Turning to $a_2$, Fig.~\ref{fig-MDC-a2} shows the dependence of the MDC line shape on $a_2$.  The two constraints---the upper bound and the lower bound of $a_1$ for a given $a_2$ value, and vice versa---discussed above mean that the range of $a_2$ shown here is nearly the full range of $a_2$ allowed (panel b), or a significant fraction of it (panel a), for each given value of $a_1$.  These line shape changes are of the same kind of line shape change that is controlled by $a_1$---the change in the MDC asymmetry.  Instead of using two parameters to control the same aspect of the data, we found it sufficient to fix the value of $a_2$ at 2, and use $a_1$ as a fit parameter to describe the MDC asymmetry.  In particular, note that in panel b, which corresponds to our final $a_1$ value, the line shape depends little on $a_2$.


\end{document}